\documentclass{midl} 

\usepackage{newtxtext}
\usepackage{newtxmath}
\usepackage{cleveref}
\usepackage{tikz}
\usetikzlibrary{shapes}
\usetikzlibrary{positioning}
\usepackage{booktabs}

\usepackage{mwe} 

\jmlrproceedings{MIDL}{Medical Imaging with Deep Learning}
\jmlrpages{}
\jmlryear{2019}
\jmlrworkshop{MIDL 2019 -- Extended Abstract Track}

\title{Shadow Detection for Ultrasound Images Using Unlabeled Data and Synthetic Shadows}

\midlauthor{%
    \Name{Suguru Yasutomi\nametag{$^{1,2}$}} \Email{yasutomi.suguru@fujitsu.com}\\
    \Name{Tatsuya Arakaki\nametag{$^{3}$}} \Email{arakakit@med.showa-u.ac.jp}\\
    \Name{Ryuji Hamamoto\nametag{$^{4,5}$}} \Email{rhamamot@ncc.go.jp}\\
    \addr $^{1}$ Artificial Intelligence Laboratory, Fujitsu Laboratories Ltd.\\
    \addr $^{2}$ RIKEN AIP-Fujitsu Collaboration Center, RIKEN\\
    \addr $^{3}$ Department of Obstetrics and Gynecology, Showa University School of Medicine\\
    \addr $^{4}$ Cancer Translational Research Team, AIP Center, RIKEN\\
    \addr $^{5}$ Division of Molecular Modification and Cancer Biology, National Cancer Center Research Institute
}

\begin{document}

\setlength{\abovedisplayskip}{2pt} 
\setlength{\belowdisplayskip}{2pt} 

\maketitle



\section{Introduction}
Medical ultrasound is a popular image diagnosing technique.
The advantage of ultrasound imaging is low introduction cost and high temporal
resolution.
On the other hand, its spatial resolution tends to be low, and
this leads clinicians to overlook small lesions.
To alleviate this, there have been many attempts to support diagnosing
by image recognition~\cite{Noble2006,Cheng2010}.
However, ultrasound images often suffer from shadows caused mainly by bones.
The shadows prevent not only diagnosing accurately from clinicians but also
working properly from the image recognition methods.
Detecting such shadows can be the first step to deal with them.
Once shadows detected, we can screen data for image recognition as preprocessing,
alert clinicians to low-quality images,
and so forth.
Shadow detection methods had been performed using traditional
image processing~\cite{Hellier2010,Karamalis2012}.
Recently, deep learning based methods have been proposed~\cite{Meng2018,Meng2018a}.
The traditional methods rely on domain-specific knowledge,
and thus it is costly to apply to multiple different domains.
The deep learning based methods learn optimal feature extraction
automatically from training data,
but they need segmentation labels which are expensive.

In this paper, we propose a novel shadow detection method
that can be learned using only unlabeled data
by utilizing the feature extraction capability of deep learning and
relatively coarse domain-specific knowledge.
Namely, our method is based on a restricted auto-encoding~\cite{Vincent2010} structure
and synthetic shadows.
We construct the structure that separates input images into
shadows and other contents and then combines them to reconstruct input images.
To encourage the network to separate the input into these images,
it is learned to predict synthetic shadows that are injected to the input beforehand.
We show that our method can detect shadows effectively by experiments on
ultrasound images of fetal heart diagnosis.

\section{Proposed method}
\subsection{Network Structure}
\begin{figure}[t]
    \centering
    \scalebox{0.87}{%
        \small
        \begin{tikzpicture}
            \tikzset{static/.style={draw, rectangle, rounded corners, align=center, fill=blue!10}};
            \tikzset{op/.style={draw, rectangle, fill=green!10}};
            \tikzset{enc/.style={draw, align=center, text width=1.6cm, trapezium, trapezium angle=80, shape border rotate=270, fill=green!10}};
            \tikzset{dec/.style={draw, align=center, text width=1.6cm, trapezium, trapezium angle=80, shape border rotate=90, fill=green!10}};

            \node[static, text width=3cm] (input) at (0, -1.2) {\includegraphics[width=1.5cm]{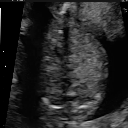}\\Input image $x$};
            \node[static, text width=3cm] (ashadow) at (0, 1.2) {\includegraphics[width=1.5cm]{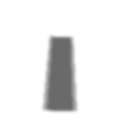}\\Synthetic shadow $x_s$};
            \node[op] (add-shadow) at (2.15, 0) {$\times$};
            \node[static, text width=2cm] (input-ashadow) at (3.7, 0) {\includegraphics[width=1.5cm]{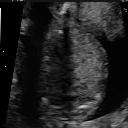}\\Input with synthetic shadow $\tilde{x}$};
            \node[enc] (encoder) at (6, 0) {Encoder\\$E$};
            \node[dec] (shadow-decoder) at (8, 1.2) {Shadow decoder\\$D_s$};
            \node[dec] (content-decoder) at (8, -1.2) {Content decoder\\$D_c$};
            \node[static, text width=2.6cm] (shadow) at (10.7, 1.2) {\includegraphics[width=1.5cm]{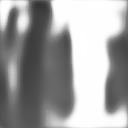}\\Shadow image $\hat{x}_s$};
            \node[static, text width=2.6cm] (content) at (10.7, -1.2) {\includegraphics[width=1.5cm]{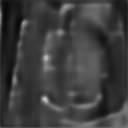}\\Content image $\hat{x}_c$};
            \node[op] (add-shadow-pred) at (12.6, 0) {$\times$};
            \node[static] (reconstruction) at (14.4, 0) {\includegraphics[width=1.5cm]{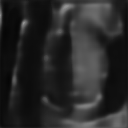}\\Reconstruction $\hat{x}$};
            \node[static] (beta) at (10.7, -3) {Beta distribution};

            \draw[-latex] (input) -- (add-shadow);
            \draw[-latex] (ashadow) -- (add-shadow);
            \draw[-latex] (add-shadow) -- (input-ashadow);
            \draw[-latex] (input-ashadow) -- (encoder);
            \draw[-latex] (encoder) -- (shadow-decoder);
            \draw[-latex] (encoder) -- (content-decoder);
            \draw[-latex] (shadow-decoder) -- (shadow);
            \draw[-latex] (content-decoder) -- (content);
            \draw[-latex] (shadow) -- (add-shadow-pred);
            \draw[-latex] (content) -- (add-shadow-pred);
            \draw[-latex] (add-shadow-pred) -- (reconstruction);

            \draw[latex-latex, dotted] (ashadow.north) -- ++(0, 0.5) -| node[xshift=-7cm, below]{MSE only for synthesized shadow area} (shadow.north);
            \draw[latex-latex, dotted] (input-ashadow.south) -- ++(0, -2) -| node[xshift=-7cm, above]{MSE} (reconstruction.south);
            \draw[latex-latex, dotted] (content) -- node[left]{NLL} (beta);
        \end{tikzpicture}
        }
    \caption{%
        Overview of the proposed method.
        Solid arrows and dotted arrows represent data flow and calculation of losses, respectively.}
    \label{fig:overview}
\end{figure}
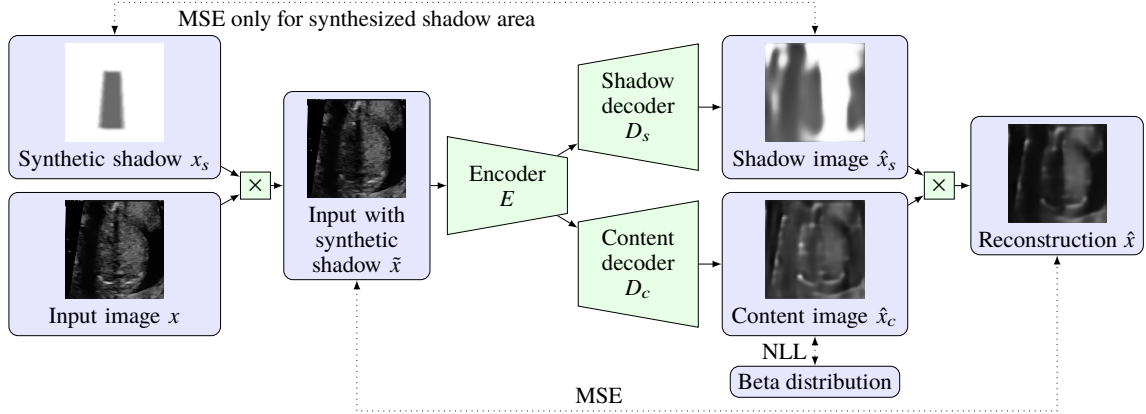

\Cref{fig:overview} shows the overview of the proposed method.
The network archtecture is based on auto-encoders, but it consists of
1) single encoder $E$ that extracts feature $z$,
2) shadow decoder $D_s$ that predicts shadow images $\hat{x}_s$,
3) content decoder $D_c$ that predicts content without shadows $\hat{x}_c$, and
4) generation of reconstruction images $\hat{x}$ by element-wise product of $\hat{x}_s$ and $\hat{x}_s$.
The encoder $E$ is a ConvNet, and the decoders $D_s$ and $D_c$ are DeconvNets.
The features and images are defined as follows:
\begin{equation*}
    z = E(x),\ \hat{x}_s = \mathrm{sigmoid}(D_s(z)),\ \hat{x}_c = \mathrm{sigmoid}(D_c(z)),\ \hat{x} = \hat{x}_s \circ \hat{x}_c,
\end{equation*}
where $x \in [0, 1]^{W\times H}$ is the input image and $\circ$ denotes Hadamard product.
The loss function for the auto-encoding structure is given as mean squared error (MSE),
i.e. $l_{\mathrm{AE}}(x, \hat{x}) = \frac{1}{WH}\sum_{ij} (\hat{x}_{ij} - x_{ij})^2$.

\subsection{Synthetic Shadow Injection and Prediction}
The normal learning process for auto-encoders does not lead $\hat{x}_s$ and $\hat{x}_c$ to contain
only shadows and only contents, respectively.
To make $D_s$ output only shadows, we inject synthetic shadows $x_s$ to input images and
lead $D_s$ to predict them.
As we know which kind of shadows may appear depending on the type of probes,
it is relatively easy to inject plausible synthetic shadows.
In this work, we focus on convex probes and generate shadows as random annular sectors
in a rule-based manner.
The injection of the synthetic shadows $x_s \in [0, 1]^{W\times H}$ can be written as
$\tilde{x} = x \circ x_s$,
and we replace the input to the network from $x$ to $\tilde{x}$.
Regarding that the background of $x_s$ is $1$ and the synthetic shadows are
expressed as a range of $[0, 1)$,
we define the loss function for shadow prediction as
\begin{equation*}
    l_s(x_s, \hat{x}_s) = \frac{1}{WH}\sum_{ij} \mathbf{1}[\hat{x}_{s_{ij}}\neq 1](\hat{x}_{s_{ij}} - x_{s_{ij}})^2.
\end{equation*}
It evaluates the correctness of predicted shadows w.r.t.\ the area that the synthetic shadows exist
because we do not know whether shadows exist on other area.

\subsection{Loss function}
Besides loss functions mentioned above, we use two additional functions.
One is regularization for predicted shadows $\hat{x}_s$ that prevent them from getting too dark:
$l_\mathrm{sreg}(\hat{s}) = \frac{1}{WH}\sum_{ij} \left| 1 - \hat{x}_{s_{ij}} \right|$.
The other is to restrict the predicted content $\hat{x}_c$ to distributed under
some beta distribution~\cite{Bishop2006}: $l_c(\hat{x}_c) = -\sum_{ij}\ln \left[p_{\mathrm{beta}}(\hat{x}_{c_{ij}} | \alpha, \beta)\right]$,
i.e.\ negative log likelihood (NLL) under the distribution.
The resulting loss function is given as
\begin{equation*}
    l =
    \lambda_{\mathrm{AE}}l_{\mathrm{AE}}(\tilde{x}, \hat{x})
    + \lambda_{s}l_{s}(x_s, \hat{x}_s) \nonumber\\
    + \lambda_{\mathrm{sreg}}l_{\mathrm{sreg}}(\hat{x}_s)
    + \lambda_{c}l_{c}(\hat{x}_c),
\end{equation*}
where $\lambda_{\mathrm{recon}},\lambda_{\mathrm{shadow}},\lambda_{\mathrm{sreg}}$, and $\lambda_{\mathrm{content}}$ are weights for the losses.

\section{Experiments}
\begin{figure}[t]
    \centering
    \includegraphics[width=0.48\linewidth]{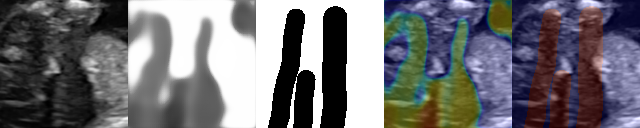} 
    \includegraphics[width=0.48\linewidth]{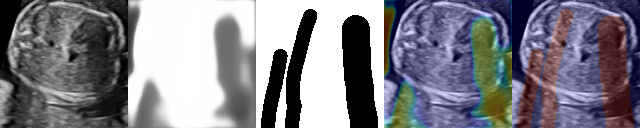}\\\vspace{0.5mm}
    \includegraphics[width=0.48\linewidth]{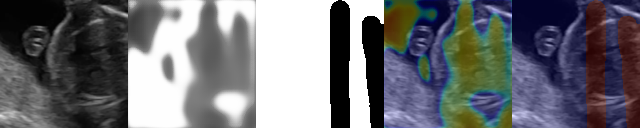} 
    \includegraphics[width=0.48\linewidth]{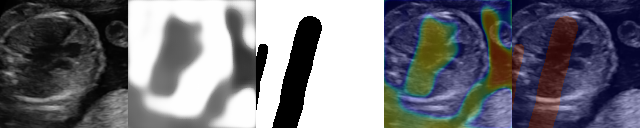}
    \caption{Example of shadow detection. From left to right; input image, shadow prediction, ground truth, shadow prediction overlayed to input, and ground truth overlayed to input.}
    \label{fig:result-examples}
\end{figure}

We evaluated the performance of the proposed method using ultrasound images of
fetal heart diagnosis.
Data for the experiments were acquired in
Showa University Hospital,
Showa University Toyosu Hospital,
Showa University Fujigaoka Hospital, and
Showa University Northern Yokohama Hospital.
All the experiments were conducted in accordance with the ethical committee of
each hospital.
Dataset consisted of 107 videos of 107 women who are 18--34 weeks pregnant.
The dataset was split into training data, validation data, and test data.
Training data was 93 videos which were converted to 37378 images.
The validation data was 61 images picked up from 7 videos.
The test data was 52 images picked up from 7 videos.
The validation data and the test data were pixel-level annotated by clinicians.
We compared the performance of the proposed method with trivial image thresholding
as a baseline, and with SegNet~\cite{Badrinarayanan2017} as a reference of
simple deep learning approach.
Because SegNet is a supervised method, we trained it with a part of the validation data.
Five videos in the validation data were used as training data for SegNet,
leaving two videos for validation.
For all methods, the hyperparameters were selected using the validation data.
\Cref{tab:results} shows the results.
\begin{table}[t]
    \centering
    \small
    \caption{Experimental results. Each score is average over images.}
    \label{tab:results}
    \begin{tabular}{lll}
        \toprule
        Method               & IoU                           & DICE \\\midrule
        Trivial thresholding & $0.229\ (\pm 0.105)$          & $0.361\ (\pm 0.140)$ \\
        SegNet               & $0.338\ (\pm 0.150)$          & $0.486\ (\pm 0.179)$ \\
        Proposed method      & $\mathbf{0.340}\ (\pm 0.132)$ & $\mathbf{0.492}\ (\pm 0.161)$ \\\bottomrule
    \end{tabular}
\end{table}
It shows that our method fairly more effective than the trivial one.
Moreover, the proposed method achieved slightly better performance than SegNet.
This result indicates that our method works well in situations with
small annotated data.
\Cref{fig:result-examples} shows examples of shadow detection with our method.
They illustrate that our method can detect various shadows,
although it tends to predict dark areas (e.g.\ amniotic fluid and cardiac cavity) as shadows.

\section{Conclusion}
In this paper, we proposed a shadow detection method for ultrasound images
that can be learned by unlabeled data.
By experimental results, the effectiveness of the method is shown.
Since the method uses only unlabeled data, it can be easy to apply to
multiple different domains (e.g.\ different machines and different organs).
Evaluations in such situations would be future work.

\midlacknowledgments{We would like to thank Akira Sakai, Masaaki Komatsu, Ryu Matsuoka, Reina Komatsu, Mayumi Tokunaka, Hidenori Machino, Kanto Shozu, Ai Dozen, Ken Asada, Syuzo Kaneko, and Akihiko Sekizawa for useful discussions and data acquisition.}

\bibliography{library}


\end{document}